\documentclass[twocolumn,prb]{revtex4}
\newcommand{\lsim} 
 {\ \raise.35ex\hbox{$<$}\kern-0.75em\lower.5ex\hbox{$\sim$}\ }
\newcommand{\gsim}
 {\ \raise.35ex\hbox{$>$}\kern-0.75em\lower.5ex\hbox{$\sim$}\ }
\def\journal #1#2#3#4{#1 {\bf #2}, #3 (#4)}

\def\PRB{Phys.\ Rev.\ B}
\def\PRL{Phys.\ Rev.\ Lett.}

\def\JPSJ{J.\ Phys.\ Soc.\ Jpn.}

\def\PTP{Prog.\ Theor.\ Phys.}

\def\COMP{Comments~Cond.~Mat.~Phys.}
\usepackage{graphicx}
\usepackage{dcolumn}
\usepackage{bm}
\usepackage{amsmath}
\usepackage{times}
\usepackage[usenames]{color}
\usepackage{ulem}

\begin{document}

\title{Band and Mott Insulators and Superconductivity in Honeycomb-Lattice Ionic-Hubbard Model}
\author{ Tsutomu Watanabe$^{1}$ and Sumio Ishihara$^{2,3}$ }
\affiliation{$^1$Department of Natural Science, Chiba Institute of Technology, Narashino 275-0023, Japan \\
$^2$ Department of Physics, Tohoku University, Sendai 980-8578, Japan \\
$^3$ Core Research for Evolutional Science and Technology (CREST), Sendai 980-8578, Japan \\
}
\date{\today}
\begin{abstract}
Motivated by superconductivity (SC) in layered nitrides, we study an ionic-Hubbard model on a honeycomb lattice, which consists of two sublattices with an energy-level offset, by using an optimization variational Monte Carlo method. Parameter values for the Coulomb interaction and 
the energy-level offset, to realize a band insulator in non-doped state, 
are evaluated, and the SC state is investigated 
in a carrier-doped band insulator. 
It is found that, in a weakly doped band insulator, spin fluctuation remains and the spin-singlet $d$-wave SC state is realized. 
Present results support an unconventional spin-singlet SC state suggested by the experimental observations in layered nitrides. 
\end{abstract}

\maketitle



%
%

%



\section{Introduction\label{sec:intro}}

Most of the superconductivity with high-$T_{\rm c}$ occur in layered materials, such as cuprates,\cite{Bednorz} low-dimensional organic conductors,
\cite{kappa1,kappa2,kappa3} and Fe-based pnictides.\cite{Kamihara} 
Because the high-$T_{\rm c}$ superconductivity is often induced close to the magnetic ordered phase, 
spin fluctuation is considered to be closely related to the Cooper pairing. 
Nitride $\beta$-$M$NCl ($M$ =Hf, Zr) doped with carriers is recognized as one of the layered superconductor with relatively high $T_{\rm c}$, which is around 
15K for Zr and 26K for Hf.\cite{Yamanaka1,Yamanaka2} 
In a series of materials, the electron conduction occurs in $M$N bilayers 
with honeycomb lattice structure, 
and the superconductivity is induced by doping of electrons into the bilayers, 
namely by the alkali-metal intercalation.\cite{Hase,Felser,Weht} 
In contrast to most of the high-$T_{\rm c}$ layered superconductivities, 
this nitride is a band insulator without a magnetic order. 

Experimental researches on the layered nitride superconductors, Li$_x$$M$NCl, have revealed many interesting features in the superconducting (SC) state. 
Nuclear magnetic resonance (NMR) \cite{Tou} and muon spin relaxation measurements \cite{Uemura,Ito} have suggested the quasi two-dimensional superconductors. 
Regardless of the relatively high $T_{\rm c}$, the $T$-linear specific heat has been found to be very small ($\sim$ 1mJ/mol$^2$).\cite{Taguchi} 
The isotope effect is weak \cite{Tou2,Taguchi2} and the SC gap ratio 2$\Delta/k_{\rm B}T_{\rm c}$ is around $5$ for Li$_{0.12}$ZrNCl \cite{Taguchi} and Li$_{0.48}$(THF)$_3$HfNCl,\cite{Ekino} which is unexpectedly high in comparison with the BCS theory. 
The NMR~\cite{Tou2,Hotehama} and specific heat measurements \cite{Kasahara} suggest that the SC gap is of an unconventional spin singlet pairing and anisotropic.
These results show that the SC in Li$_x$$M$NCl is not of a conventional BCS type. 
A theoretical calculation for the SC state in Li$_x$$M$NCl was done in the two-band honeycomb lattice model where the fluctuation exchange (FLEX) method was applied.\cite{Kuroki} 
The $d+id'$-type SC gap was suggested to be stabilized 
in the spin fluctuation mechanism. 

We study the ionic-Hubbard model on a honeycomb lattice, which consists of two sublattices with an energy-level offset, with the conducting plane of Li$_x$$M$NCl in mind.  
We adopt an optimization variational Monte Carlo (VMC) method. 
Beyond the previous theoretical approaches, 
this method enables us to accurately evaluate Mott and band insulators and a metal to insulator transition, despite the coupling strength of electron-electron interaction. 
First, in the non-doped system, we determine the parameter values for the Coulomb interaction and the energy-level offset to realize the band insulator. 
Next, in the system corresponding to the doped band insulator, we investigate stabilities of the SC states with several pairing symmetries and discuss a  mechanism of the paring. 
Spin fluctuation survives with sufficient magnitude even for large level offsets, for which a band insulator is realized at half-filling. 
We find that the SC with the $d$-wave pairing symmetry is stabilized in the doped band insulating system. 

In Sect.~\ref{sec:method}, the ionic-Hubbard model on a honeycomb lattice and the formulation of the VMC method are introduced. 
In Sect.~\ref{sec:nondope}, the ground-state phase diagram in the non-doped system is presented, and the parameter region in which the band insulator is realized is determined. 
In Sect.~\ref{sec:super}, the numerical results for the SC state in a carrier-doped band insulator are shown. 
Section \ref{sec:summary} is devoted to the summary and discussion. 
A part of the present results was briefly reported in Ref.~21. 

\section{Model and method \label{sec:method}}

As an effective model for Li$_x$$M$NCl, we consider a Hubbard model on a single-layer honeycomb lattice consisting of alternating g$M$h and gNhsites with an energy-level offset. The Hamiltonian is given as follows, 
\begin{eqnarray}
{\cal H} = {\cal H}_t + {\cal H}_{t'} + {\cal H}_\Delta + {\cal H}_U , 
\label{eq:Ham}
\end{eqnarray}
with 
\begin{eqnarray}
{\cal H}_t = t\sum\limits_{<i,j>\sigma } {(c_{i\sigma }^{{\rm{A}}\dag }c_{j\sigma }^{\rm{B}} + {\rm{H}}{\rm{.c}}{\rm{.}})}, 
\label{eq:Ht}
\end{eqnarray}
\begin{eqnarray}
{\cal H}_{t'} = t'\sum\limits_{(i,j)\sigma } {(c_{i\sigma }^{{\rm B}\dag }c_{j\sigma }^{\rm{B}} + {\rm{H}}{\rm{.c}}{\rm{.}})}, 
\label{eq:Ht'}
\end{eqnarray}
\begin{eqnarray}
{\cal H}_\Delta = - \Delta \sum\limits_{i} {n^{\rm A}_i} + \Delta \sum\limits_{i} {n^{\rm B}_i}, 
\label{eq:HD}
\end{eqnarray}
and 
\begin{eqnarray}
{\cal H}_U = U \sum\limits_i { (n^{\rm A}_{i \uparrow} n^{\rm A}_{i \downarrow} + n^{\rm B}_{i \uparrow} n^{\rm B}_{i \downarrow}) }. 
\label{eq:HU}
\end{eqnarray}
Here, $t$ is the electron hopping from a site in sublattice A(B) to the nearest-neighbor (NN) sites in sublattice B(A), $t'$ is the electron hopping from a site in sublattice B to the NN site in sublattice B. 
We define the number operatore $n^\lambda_i = n^\lambda_{i\uparrow} + n^\lambda_{i\downarrow}$ with $n^\lambda_{i\sigma} = c^{\lambda \dag}_{i\sigma} c^\lambda_{i\sigma}$ where $\lambda$(=A, B) is an index for the sublattices. 
The level offset for the two orbitals is denoted by 2$\Delta$. 
We do not introduce the hopping from a site in sublattice A to the 
NN site in sublattice A. 
This is supported by the tight-binding fit of the energy bands calculated by the first-principle calculation.\cite{Kuroki,Weht} 

\begin{figure}[t]
\begin{center}
\includegraphics[width=7.5cm,height=3.5cm]{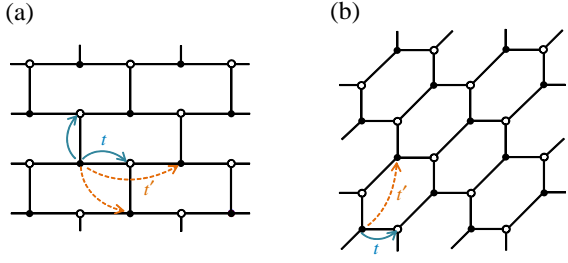}
\end{center}
\vspace{-0.3cm}
\caption{(Color online)
Two types of honeycomb lattice models: (a) Type I and (b) Type II. 
}
\label{fig:hon}
\end{figure}

For the following technical reasons, we adopt the two types of the honeycomb lattice model, termed Type I and Type II, as shown in Fig.~\ref{fig:hon}. 
Type I is the so-called "brick model", in which a staggered order is characterized by the momentum ($\pi,\pi$). This model is suitable to investigate the spin and charge correlations near the metal-to-insulator transition.  
Type II is useful to examine the SC pairing symmetries in a carrier-doped system because of the trigonal 
symmetry in the lattice structure. 
Results are independent of the types of the model.

We utilize an optimization VMC method, which can correctly treat the local 
electron correlation in the whole parameter space spanned by $U$ and $\Delta$. 
As a variational wave function, we use the Jastrow type, $\Psi = P\Phi$, where $\Phi$ is an one-body (Hartree-Fock) part, and ${\cal P}$ is a many-body correlation factor. 
As for ${\cal P}$, in addition to the well-knowm Gutzwiller (onsite) factor give by 
\begin{eqnarray}
{\cal P}_G  = \prod\limits_{i\lambda} 
{\left[ {1 - (1 - g)n^\lambda_{i \uparrow } n^\lambda_{i \downarrow } } \right]}, 
\label{eq:PG}
\end{eqnarray}
we introduce an intersite correlation factor ${\cal P}_{\rm Q}$ given by 
\begin{eqnarray}
{\cal P}_Q  = \prod\limits_{i\lambda} 
{\left( {1 - \mu Q_i^{\lambda}} \right)}  , 
\label{eq:PQ}
\end{eqnarray}
with 
\begin{eqnarray}
Q_i^{\lambda} = \prod\limits_\tau 
{\left[ {d^\lambda_i \left( {1 - e^{\lambda'}_{i + \tau} } \right) 
+ e^\lambda_i \left( {1 - d^{\lambda'}_{i + \tau} } \right)} \right]}. 
\label{eq:Qi}
\end{eqnarray}
Here, $d^\lambda_i = n^\lambda_{i \uparrow} n^\lambda_{i \downarrow}$, 
$e^\lambda_i = (1 - n^\lambda_{i \uparrow})(1 - n^\lambda_{i \downarrow})$, $g$ and $\mu$ are the variational parameters, 
$\tau$ runs over all the adjacent sites in the NN bond directions, 
and $\lambda' = ({\rm A,B})$ for $\lambda = ({\rm B,A})$. 
After all, the trial wave function is given by  
\begin{eqnarray}
\Psi = {\cal P}_Q {\cal P}_G \Phi. 
\label{eq:Psi}
\end{eqnarray}
It is known that, the binding effect of a doubly-occupied site (doublon) to an empty site (holon) is indispensable to describe the Mott insulator as well as the SC state, appropriately.\cite{Yokoyama1,Yokoyama2,Yokoyama3} 

Regarding $\Phi$, we consider two trial states: (i) a state which has an antiferromagnetic (AF) order in the Mott insulator and a charge disproportion (CD) between the A and B sublattices in the band insulator, denoted by $\Phi_{\rm ins}$, and (ii) a BCS state $\Phi_{\rm BCS}$.  
The first trial state, $\Phi_{\rm ins}$, is given by diagonalizing the Hartree-Fock (HF) Hamiltonian: 
\begin{eqnarray}
{\cal H}_{\rm ins} = 
\sum\limits_{{\bf k}\sigma } {\phi_{{\bf k}\sigma }^\dag h_{{\bf k}\sigma} \phi _{{\bf k}\sigma} }, 
\label{eq:HAFCO}
\end{eqnarray}
where 
\begin{eqnarray}
\phi _{{\bf k}\sigma } = \left( {\begin{array}{*{20}c}
   {c^{\rm A}_{{\bf k}\sigma } }  \\
   {c^{\rm B}_{{\bf k}\sigma } }  \\
\end{array}} \right), 
\label{eq:Phiks}
\end{eqnarray}
and 
\begin{eqnarray}
h_{{\bf k}\sigma }  = \left( {\begin{array}{*{20}c}
   {W_\sigma  } & {A^\dag_{\bf k} }  \\
   {A_{\bf k} } & {B_{\bf k} - W_\sigma  }  \\
\end{array}} \right). 
\label{eq:hks}
\end{eqnarray}
The operator $c^{\lambda}_{{\bf k}\sigma}$ is the Fourier transform of $c^{\lambda}_{i\sigma}$. 
The summation $\sum_{\bf k}$ runs over the reduced two-dimensional Brillouin zone. 
We define 
\begin{eqnarray}
W_\sigma = - \Delta + \Delta_{\rm CD}  + {\rm sgn}(\sigma) \Delta_{\rm AF}, 
\label{eq:Ws}
\end{eqnarray}
\begin{eqnarray}
A_{\bf k} = t\sum\limits_{\bf \tau} {e^{ - i{\bf k} \cdot {\bf \tau}} },
\label{eq:Ak}
\end{eqnarray}
and 
\begin{eqnarray}
B_{\bf k} = 2t'\sum\limits_{{\bf \tau'}} {\cos ({\bf k} \cdot {\bf \tau'})} 
\label{eq:Bk}, 
\end{eqnarray}
where $\sum_{\bf \tau}$ ($\sum_{\bf \tau'}$) represents a summation for the vectors connecting NN sites in the A-B (A-A) sublattices.  
We introduce $\Delta_{\rm AF}$ and $\Delta_{\rm CD}$ as mean fields (MF) for the AF and CD states, respectively. 
The MF electron densities are defined as 
$\left\langle {n_{i\sigma }^{\rm A}} \right\rangle = N + \Delta_{{\rm CD}} + {\mathop{\rm sgn}} (\sigma )\Delta_{{\rm AF}}$ 
and 
$\left\langle {n_{i\sigma }^{\rm B}} \right\rangle = N - \Delta_{{\rm CD}} - {\mathop{\rm sgn}} (\sigma )\Delta_{{\rm AF}}$, where $N$ is an averaged electron number per site.
We note that $\Delta_{\rm AF}$ controls the staggered spin density in an AF ordered state, and $\Delta_{\rm CD}$ controls the charge density in the charge-rich A and charge-poor B sites. 

\begin{table*}[t]
\begin{center}
\caption{\label{tab:gap} 
Gap functions $g({\bf k})$ for various pairing symmetries.
The wave number vector is defined for the honeycomb lattice model of Type II (Fig.~\ref{fig:hon}). 
}
\vspace{+0.3cm}
\begin{tabular}{c|c|c} \hline
 Symmetry & $m$ & $g({\bf k})$ \\
\hline
 $s$ & 2 & $g_{\rm BCS}$ \\
\hline
 $d_1$ & 1 & $g_{\rm BCS}(\cos k_x-\cos k_y)$ \\
\hline
 $d_2$ & 2 & $g_{\rm BCS}[\cos (2k_x+k_y) - \cos (k_x+2k_y)]$ \\
\hline
 $d$+$id$ & 1 & $g_{\rm BCS}[\cos k_x + e^{i\frac{{2\pi}}{3}}\cos(k_x+k_y) 
+ e^{i\frac{{4\pi}}{3}}\cos k_y]$ \\
\hline
 $p$ & 1 & $g_{\rm BCS}\sin k_x$ \\
\hline
 $f_1$ & 1 & $g_{\rm BCS}[\sin k_x - \sin(k_x+k_y) + \sin k_y]$ \\
\hline
 $f_2$ & 2 & $g_{\rm BCS}[\sin (2k_x+k_y) - \sin(k_x+2k_y) - \sin (k_x-k_y)]$ \\
\hline
\end{tabular}
\end{center}
\end{table*}

\begin{figure*}[t]
\begin{center}
\includegraphics[width=12.5cm,height=7.5cm]{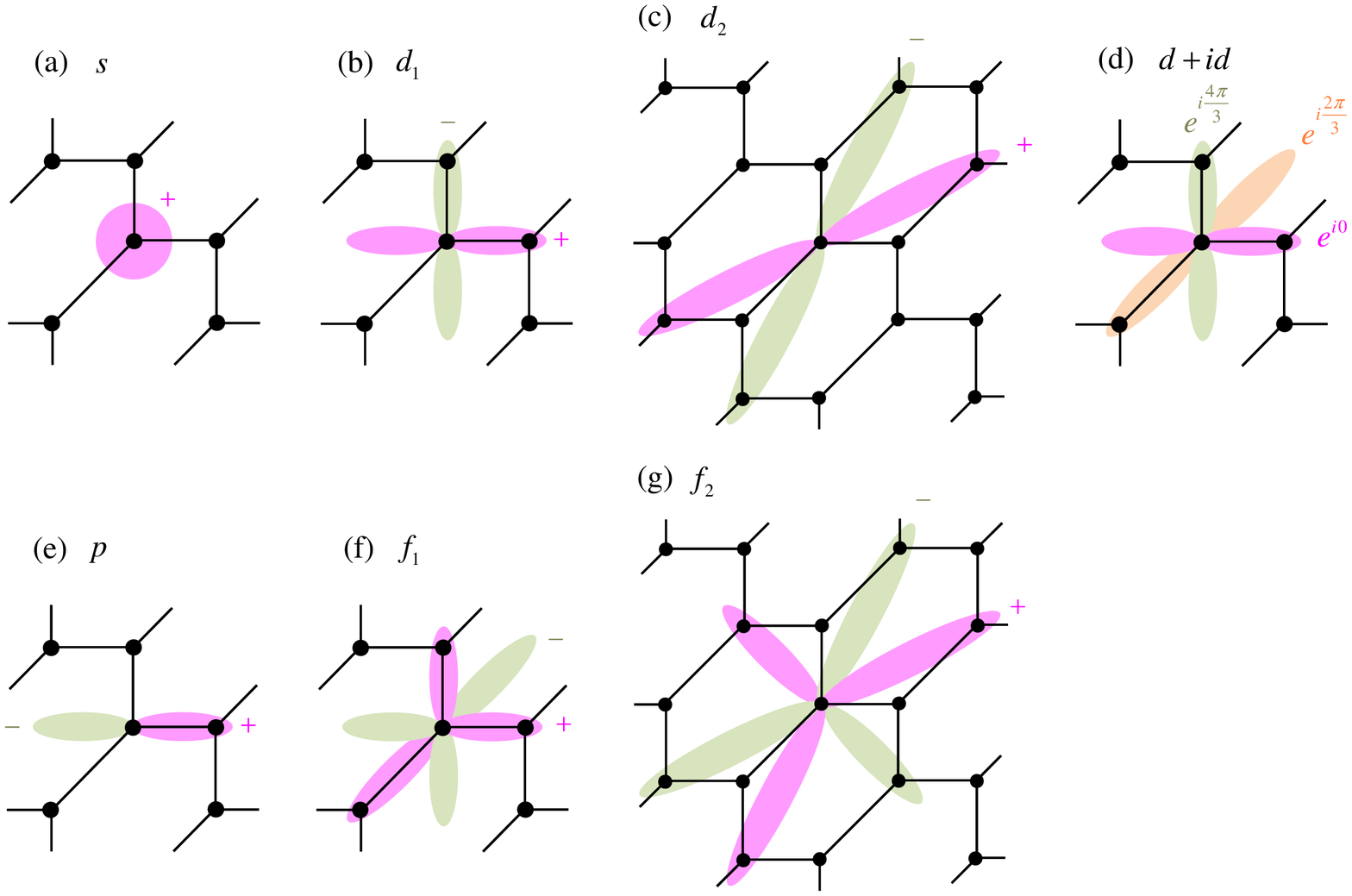}
\end{center}
\vspace{-0.3cm}
\caption{(Color online)
Schematic gap functions of (a) $s$, (b) $d_1$, (c) $d_2$, (d) $d$+$id$, (e) $p$, (f) $f_1$ and (g) $f_2$ waves.
The honeycomb lattice model of Type II (Fig.~\ref{fig:hon}) is adopted. 
}
\label{fig:gap}
\end{figure*}

In order to construct the BCS state $\Phi_{\rm BCS}$, we introduce the following HF Hamiltonian: 
\begin{eqnarray}
{\cal H_{{\rm{BCS}}}} = {\cal H}_{\rm K} + {\cal H}^{(m)}_{\rm pair}
\label{eq:HBCS}, 
\end{eqnarray}
where 
\begin{eqnarray}
{\cal H}_{\rm K} = \sum\limits_{{\bf{k}}\sigma } {\phi_{{\bf{k}}\sigma}^\dag {{h'}_{{\bf{k}}\sigma}}{\phi_{{\bf{k}}\sigma }}} , 
\label{eq:tDel} 
\end{eqnarray}
and 
\begin{eqnarray}
{h'_{{\bf{k}}\sigma }} = \left( {\begin{array}{*{20}{c}}
{ - \Delta + \Delta_{\rm CD} - \mu_0 }&{A_{\bf{k}}^\dag }\\
{{A_{\bf{k}}}}&{{B_{\bf{k}}} + \Delta - \Delta_{\rm CD} - \mu_0 }
\end{array}} \right)
\label{eq:hp}. 
\end{eqnarray}
The pairing term ${\cal H}^{(m)}_{\rm pair}$ is defined by 
\begin{eqnarray}
{\cal H}^{(1)}_{\rm{pair}} = \sum\limits_{\bf{k}} {g({\bf{k}})(c_{{\bf{k}} \uparrow }^{\rm{A}}c_{ - {\bf{k}} \downarrow }^{\rm{B}} + c_{{\bf{k}} \uparrow }^{\rm{B}}c_{ - {\bf{k}} \downarrow }^{\rm{A}})} + {\rm H.c.} 
\label{eq:Hpa1} ,
\end{eqnarray}
and 
\begin{eqnarray}
{\cal H}^{(2)}_{\rm{pair}} = \sum\limits_{\bf{k}} {2g({\bf{k}})(c_{{\bf{k}} \uparrow }^{\rm{A}}c_{ - {\bf{k}} \downarrow }^{\rm{A}} + c_{{\bf{k}} \uparrow }^{\rm{B}}c_{ - {\bf{k}} \downarrow }^{\rm{B}})} + {\rm H.c.}
\label{eq:Hpa2} , 
\end{eqnarray}
where $m$ takes 1 for a pair of electrons between the different sublattices (A-B), and takes 2 between the same sublattices (A-A or B-B), and $g({\bf k})$ is a gap function for the pairing. 
%
%
The chemical potential $\mu_0$ is fixed at a value determined from the tight-binding band of ${\cal H}_{\rm K}$, which depends on the optimized parameter 
$\Delta_{\rm CD}$. 
In Eq.~(\ref{eq:hp}), the MF order parameter of AF, $\Delta_{\rm AF}$, 
is not considered, 
because we are mainly interested in the SC state arising 
in the nonmagnetic doped band insulating state. 

By diagonalizing ${\cal H}_{\rm K}$, we have the two bands $E_{\bf k}^{\alpha}$ 
$(\alpha=h,l)$ and the eigen operators $e^{\alpha}_{\bf k}$. 
By representing $c_{{\bf{k}} \sigma}^{\rm{A(B)}}$ by $e^{h}_{\bf k}$ and $e^{l}_{\bf k}$, the Hamiltonian is rewritten as  
%
\begin{eqnarray}
{{\cal H}_{{\rm{BCS}}}} = \sum\limits_{{\bf{k}}\alpha \sigma } {E_{\bf{k}}^\alpha e_{{\bf{k}}\sigma }^{\alpha \dag }e_{{\bf{k}}\sigma }^\alpha } + \sum\limits_{{\bf{k}}\alpha } {D_{{\bf{k}}\alpha }^{(m)} e_{{\bf{k}} \uparrow }^\alpha e_{ - {\bf{k}} \downarrow }^\alpha } . 
\label{eq:HBCSp} 
\end{eqnarray}
We define 
\begin{eqnarray}
D^{(1)}_{{\bf k}\alpha} = \frac{{{\mathop{\rm Re}\nolimits} 
{A_{\bf{k}}}}}{{\sqrt {{{\left| {{A_{\bf{k}}}} \right|}^2} + 
{ (\Delta - \Delta_{\rm CD})^2}} }} {\rm sgn}(\alpha) g({\bf{k}})  ,
\label{eq:D1} 
\end{eqnarray}
\begin{eqnarray}
D^{(2)}_{{\bf k}\alpha} = g({\bf{k}}) , 
\label{eq:D2} 
\end{eqnarray}
and 
\begin{eqnarray}
E_{\bf k}^\alpha = \tilde \varepsilon _{\bf k}^\alpha  - \mu_0, 
\label{eq:Ek} 
\end{eqnarray}
where ${\rm sgn}(\alpha) = 1$ ($-1$) for $\alpha = h$ ($l$),\cite{pair} 
and 
\begin{eqnarray}
\tilde \varepsilon_{\bf k}^\alpha = {\rm sgn}(\alpha) 
\sqrt {\left| {A_{\bf k} } \right|^2 + 
(\Delta - \Delta_{\rm CD})^2}. 
\label{eq:epk} 
\end{eqnarray}

From this Hamiltonian, the BCS state is given by  
\begin{eqnarray}
\Phi_{{\rm BCS}} = \sum\limits_{{\bf k}\alpha} {\left( { \varphi_{\bf k}^\alpha e_{{\bf k}\uparrow}^{\alpha\dag} e_{ - {\bf k}\downarrow}^{\alpha\dag} } \right)^{\frac{{N_{\rm e}}}{2}} } \left| 0 \right\rangle
\label{eq:BCS}, 
\end{eqnarray}
in which $N_{\rm e}$ is a fixed electron number, and the BCS coherence factor is defined as 
\begin{eqnarray}
\varphi_{\bf k}^\alpha = \frac{{D^{(m)}_{{\bf k}\alpha}}}{{E_{\bf k}^\alpha + \sqrt {(E_{\bf k}^\alpha)^2 + \left| {D^{(m)}_{{\bf k}\alpha}} \right|^2 } }} . 
\label{eq:Phik} 
\end{eqnarray}

As for the gap function $g({\bf k})$, we factorize it as 
\begin{eqnarray}
g({\bf k}) = g_{{\rm BCS}} z_{\bf k}, 
\label{eq:Delk} 
\end{eqnarray}
where $z_{\bf k}$ represents anisotropy of the gap function and $g_{\rm BCS}$ is a variational parameter to be optimized. 
We study a variety of pairing symmetries: the $s$, $d_1$, $d_2$, $d$+$id$ waves as candidates for the singlet pair, and the $p$, $f_1$, $f_2$ waves for the triplet pair. 
Formulation and depiction of these gap functions are summarized in Table.~\ref{tab:gap} and Fig.~\ref{fig:gap}, respectively. 
The $d_1$-, $d$+$id$-, $p$- and $f_1$-wave functions correspond to the A-B pairing ($m=1$), while the $s$-, $d_2$- and $f_2$-wave functions correspond to the A-A or B-B pairing ($m=2$). 
The $d_1$ wave, corresponding to the $d_{x^2-y^2}$ wave known in the cuprate 
superconductors, has nodes along the $k_y=\pm k_x$ directions. 
The $d$+$id$-wave is the anisotropic and fully-opened gap function. 

In the numerical calculations, we adopt $2\times 10^{5}$-$5\times$10$^5$ samples in most of the VMC simulations.  
The fixed-sampling method is used to optimize the variational parameters.\cite{Umriger} 
The error in the energy expectation value is of the order of $10^{-4}t$. 
Cluster sizes of $96 \le N_{\rm S} \le 216$, 
where $N_{\rm S}$ is the number of site, 
are used with the periodic-boundary condition and the  antiperiodic-boundary condition. 
We take $N_{\rm S} = L^2$ with $L=10$-$14$ for Type I, 
and $N_{\rm S} = 2L^2/3$ with $L = 12$ for Type II. 

\section{Insulating state in non-doped system \label{sec:nondope}}

\begin{figure}[t]
\begin{center}
\includegraphics[width=7.0cm,height=6.0cm]{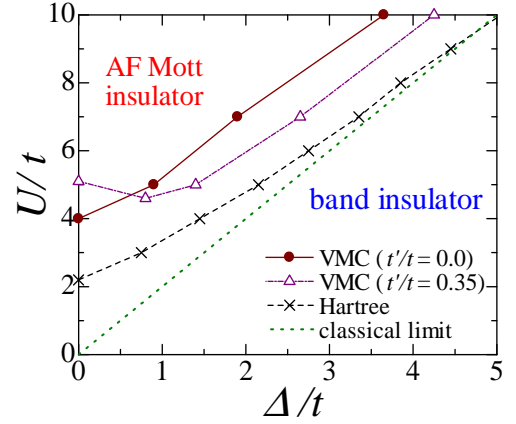}
\end{center}
\vspace{-0.5cm}
\caption{(Color online)
Phase diagram at half-filling for the most stable state of $\Psi_{\rm ins}$, in the plane of $U/t$ and $\Delta/t$. 
The boundaries between the AF Mott and band insulating regions for $t'/t = 0$ and $0.35$ are plotted by circles and triangles, respectively. 
The boundaries obtained by Hartree approximation and at classical limit 
($t=t'=0$) are also depicted by crosses and a dotted line, 
respectively. 
The system size of $N_{\rm S} = 196$ is used in the Type-I model. 
}
\label{fig:ph}
\end{figure}

We determinate the parameter set of $U$ and $\Delta$, at which a band insulator is realized in the non-doped system. 
In Fig.~\ref{fig:ph}, we show a phase diagram for the most stable state of $\Psi_{\rm ins}$ ($\equiv P\Phi_{\rm ins}$), at half-filling where 
$N_{\rm e}=N_{\rm S}$. 
In the case of $t'=0$, the phase boundary between the Mott and band insulating states is located above the boundary in  the classical limit ($t=t'=0$), 
plotted by a dotted line of $U=2\Delta$. 
This tendency of the phase boundary is also given in the Hartree approximation, and reflects the fact that energy reduction from the classical limit by quantum fluctuation is larger 
for the band insulator than that for the AF Mott insulator.  
This is because, in the AF insulator, electrons on the sublattice A have little chance to transfer to the adjacent sites, owing to the energy $2\Delta + U$. 

When $t'$ is introduced (see the case of $t'/t=0.35$ in 
Fig.~\ref{fig:ph}), the phase boundary shifts to the band-insulator side 
for $\Delta/t \gsim 1$, and shifts to the AF Mott-insulator side for 
$\Delta/t < 1$. 
The former is explained by the purturbational picture with respect to 
$t/U$ and $t'/U$; 
the energy reduction in the AF Mott insulator for $t'/t = 0.35$ is 
$24t^2t'U\Delta /(U^2-4\Delta^2)^2$ larger than that 
for $t'/t = 0$ in the case of $\Delta \sim U/2$. 
The phase-boundary shift for $\Delta/t < 1$ is explained by a suppression of the AF order due to the frustration effect caused by $t'$. 

\begin{figure}[t]
\begin{center}
\includegraphics[width=7.5cm,height=5.5cm]{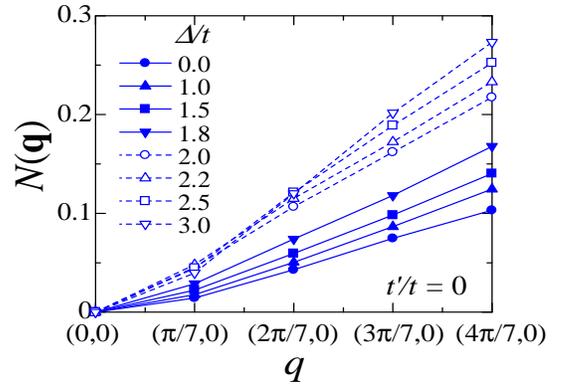}
\end{center}
\vspace{-0.5cm}
\caption{(Color online)
Charge structure factor $N({\bf q})$ in the direction of ${\bf q} = (0,0)$-$(\pi,0)$ for different values of $\Delta$ at half-filling. 
The parameters are chosen to be $U/t =7$ and $t'=0$. 
Filled symbols are for $\Delta < \Delta_{\rm c}$, and open symbols for 
$\Delta > \Delta_{\rm c}$, where $\Delta_{\rm c}$ is the boundary between the band and Mott insulating phases. 
The system size of $N_{\rm S} = 196$ is used in the Type I model. 
}
\label{fig:Nq}
\end{figure}
%
Insulating properties are directly confirmed by calculating the charge structure factor, 
\begin{eqnarray}
N({\bf q}) &=& \frac{1}{{N_{\rm S}}}\sum\limits_{ij\lambda\lambda'} 
{e^{i{\bf q} \cdot ({\bf R}^\lambda_i - {\bf R}^{\lambda'}_j)}} \nonumber\\ 
&\times& \left( {\left\langle {N^\lambda_i N^{\lambda'}_j} 
\right\rangle - \left\langle {N^\lambda_i} 
\right\rangle \left\langle {N^{\lambda'}_j} \right\rangle } \right), 
\label{eq:Nq} 
\end{eqnarray}
where 
$N^\lambda_i = \sum\nolimits_{\sigma} {n^\lambda_{i\sigma } }$. 
It is known in the variational theory that an insulating gap in the charge 
degree of freedom is indicated by the quadratic behavior in the ${\bf q}$ 
dependence of $N({\bf q})$ in the limit of 
$|{\bf q}| \rightarrow 0$.\cite{Feynman} 
In Fig.~\ref{fig:Nq}, $N({\bf q})$ at $U/t = 7$ is shown for several values 
of $\Delta$ in the direction of ${\bf q} = (0,0)$-$(\pi,0)$. 
For all the values of $\Delta$, $N({\bf q})$ near the $\Gamma$ point 
seems quadratic in $q$. 
When $\Delta$ increases, the whole magnitude of $N({\bf q})$ is 
abruptly increased across $\Delta/t \sim 1.9$. 
As shown later, 
$\Delta/t =1.9$ is a boundary between the band and Mott insulating phases. 
A metallic state is not found within the present parameter region of $U$ and $\Delta$ in Fig.~\ref{fig:ph}, except for the Dirac point at $(\Delta,U)=(0,0)$.

\begin{figure}[t]
\begin{center}
\includegraphics[width=6cm,height=9.5cm]{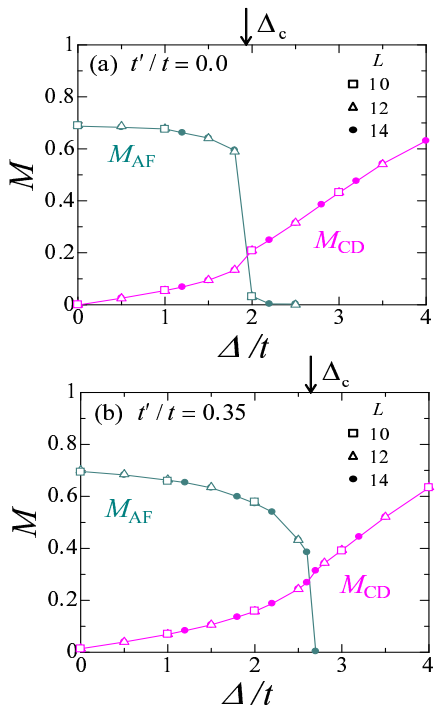}
\end{center}
\vspace{-0.5cm}
\caption{(Color online)
The order parameter for the AF Mott insulating state $M_{\rm AF}$ and that 
for the CD state $M_{\rm CD}$ for (a) $t'/t=0$ and (b) $0.35$ at half-filling. 
The parameter value is chosen to be $U/t =7$. 
The vertical arrows indicate $\Delta_{\rm c}$, the boundary between the Mott and band insulating phases. 
In both (a) and (b), the system size of $N_{\rm S} = 100$-$196$ (of $L=10$-$14$) is used in the Type I model. 
}
\label{fig:Mdw}
\end{figure}

The boundary between the AF Mott and band insulating regions is determined by 
the optimized variational parameters $\Delta_{\rm AF}$ and $\Delta_{\rm CD}$: 
we identify the AF Mott insulator for $\Delta_{\rm AF} > 0$, and 
the band insulator for $\Delta_{\rm AF} = 0$ and $\Delta_{\rm CD} > 0$. 
In order to confirm the realization of Mott and band insulators, we check 
the following two order parameters: 
\begin{eqnarray}
M_{{\rm AF}} = \frac{2}{{N_{\rm S} }}
\sum\limits_i {(S_i^{z{\rm A}} - S_i^{z{\rm B}})} , 
\label{eq:Msdw} 
\end{eqnarray}
and 
\begin{eqnarray}
M_{{\rm CD}}  = \frac{1}{{N_{\rm S}}}\sum\limits_i 
{(N^{\rm A}_i-N^{\rm B}_i)}, 
\label{eq:Msdw} 
\end{eqnarray}
where 
$S_i^{z\lambda}  = \frac{1}{2}(n^\lambda_{i \uparrow} - n^\lambda_{i \downarrow})$. 
We introduce $M_{{\rm AF}}$ as an order parameter for spin density wave in the AF Mott insulator and $M_{{\rm CD}}$ for the CD state where $\left\langle {N_i^{\rm A}} \right\rangle > 1$ and $\left\langle {N_i^{\rm B}} \right\rangle < 1$. 
In Fig.~\ref{fig:Mdw}, $M_{{\rm AF}}$ and $M_{{\rm CD}}$ are shown as functions of $\Delta$ at $U/t =7$. 
In both cases of $t'/t=0$ and $0.35$, 
$M_{{\rm AF}}$ has a large value for small $\Delta$ and vanishes at a critical value of $\Delta$ ($\Delta_{\rm c}=1.9t$). 
On the other hand, $M_{{\rm CD}}$ is always positive. 
This value gradually increases with $\Delta$ and shows large magnitude for $\Delta > \Delta_{\rm c}$, associated with a small anomaly at $\Delta_{\rm c}$. 
It is shown that the magnitudes of $M_{{\rm AF}}$ and $M_{{\rm CD}}$ are almost independent of the system size $L$. 
These results indicate the phase transition between the Mott and band insulators at $\Delta_{\rm c}$. 

\section{Superconductivity in carrier-doped system \label{sec:super}}

\begin{figure}[t]
\begin{center}
\includegraphics[width=6.5cm,height=9.0cm]{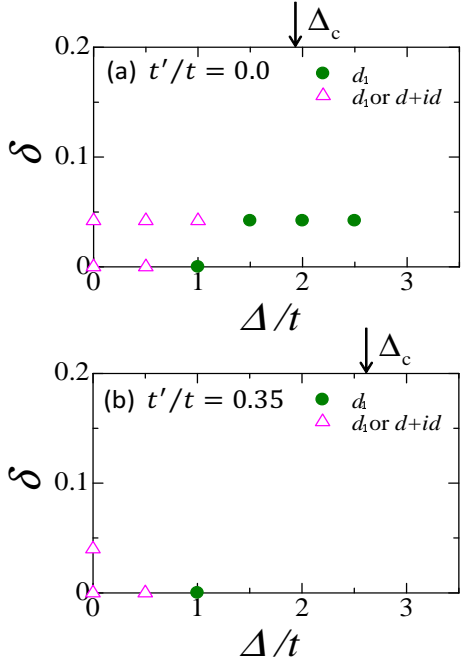}
\end{center}
\vspace{-0.3cm}
\caption{(Color online)
Phase diagram in the $\delta-\Delta/t$ plane calculated in the wave function $\Psi_{\rm BCS}$ for (a) $t'/t = 0$ and (b) $0.35$. 
Filled circles and open triangles represent the parameters where the $d_1$-wave state and both the $d_1$- and $d$+$id$-wave states are realized, respectively. 
We note that, as for the parameter spaces indicated by triangles,  
energies for the two BCS states are degenerated. 
The vertical arrows indicate $\Delta_{\rm c}$, the boundary between 
the Mott and band insulating phases. 
The parametr is chosen to be $U/t =7$, and the system size of $N_{\rm S} = 96$ is used in the Type II
model.}
\label{fig:dopeph}
\end{figure}

Now we show the results for the SC state in carrier-doped band insulators. 
Through this section, a value of $U$ is fixed at $7t$. 
As for the gap function introduced into the trial BCS state $\Psi_{\rm BCS}$ ($\equiv P\Phi_{\rm BCS}$), we check all pairing symmetries in Table.~\ref{tab:gap} for all values of $U$ and $\Delta$ studied in Fig.~\ref{fig:ph}, 
and obtain that the BCS states with the $d_1$- and $d$+$id$-wave gaps are stabilized relative to the normal state, $\Psi_{\rm BCS}(g_{\rm BCS} = 0)$. 
Therefore, we here present only the results for the $d_1$- and $d$+$id$ waves. 
In Fig.~\ref{fig:dopeph}(a), a phase diagram for the stable state of $\Psi_{\rm BCS}$ is presented in the plane of $\Delta/t$ and $\delta$, where 
$\delta(\equiv N_{\rm e}/N_{\rm S}-1)$ is the doping concentration of electrons per site. 

\begin{figure}[t]
\begin{center}
\includegraphics[width=7.0cm,height=8.0cm]{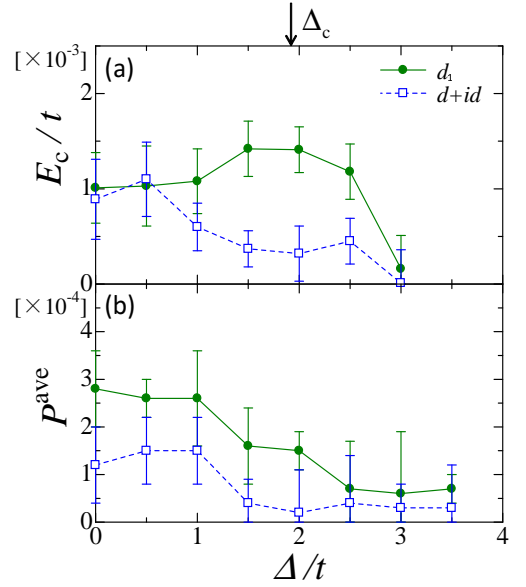}
\end{center}
\vspace{-0.5cm}
\caption{(Color online)
(a) The condensation energies $E_{\rm c}$ and (b) the averaged pair correlation function $P^{\rm ave}$ calculated in $\Psi_{\rm BCS}$ for the $d_1$- and $d$+$id$-wave gaps as functions of $\Delta/t$. 
The parameters are chosen to be $\delta = 0.04$, $t'= 0$, and $U/t =7$. 
The vertical arrow indicates $\Delta_{\rm c}$. 
The system size of $N_{\rm S} = 96$ is used in the Type II model. 
}
\label{fig:EcPd}
\end{figure}

First, we briefly touch the results at half-filling. 
The BCS states of the $d_1$- and $d$+$id$ waves are stabilized within the wave function of $\Psi_{\rm BCS}$. 
However, we note that the AF insulating state obtained from $\Psi_{\rm ins}$ is much more stable than all the BCS states in the region of $\Delta/t < 1.9$ 
for $t'/t = 0$ and $\Delta/t < 2.65$ for $t'/t = 0.35$. 
Then, we focus on the carrier-doped cases with $t'=0$. 
The $d_1$-wave state for $t'=0$ is stabilized near half-filling, namely for $\delta = 0.04$, as shown in Fig.~\ref{fig:dopeph}(a). 
We show the condensation energies for the $d_1$- and $d$+$id$-wave states at $\delta = 0.04$ in Fig.~\ref{fig:EcPd}(a). 
This is defined by $E_{\rm c} = E^{\rm Normal} - E$, in which $E$ and $E^{\rm Normal}$ are the variational energies per site 
obtained by the BCS wave function $\Psi_{\rm BCS}$ and that with $g_{\rm BCS}=0$, respectively. 
It is shown that $E_{\rm c}$ for the $d_1$-wave state has a finite value in the region of $\Delta \le 2.5t$, although the magnitude ($\sim 10^{-3}t$) is 
small relative to the condensation energy obtained in the square-lattice Hubbard model near the half-filling ($\sim 10^{-2}t$).\cite{Yokoyama3} 
These results show that in the case of $t'=0$, the $d_1$-wave state 
survives up to around $\Delta=2.5t$, 
which is larger than $\Delta_{\rm c}= 1.9t$ where the AF Mott insulator to the band insulator transition occurs at half-filling. 
On the other hand, the $d$+$id$-wave state appears only in the region of $\Delta \le 1.0t$, namely it is not stabilized in the doped band insulator. 
A realization of the $d+id$ state at $\Delta=0$ and $t'=0$ reproduces the previous numerical results.\cite{Pathak} 

In order to confirm the SC state in the case of $t'=0$, we calculate 
the pair correlation function defined as 
\begin{eqnarray}
P({\bf{r}}) = \frac{1}{{{N_{\rm{S}}}}}\sum\limits_{i\lambda} 
{\left\langle {\Delta_x^\dag ({{\bf{R}}^\lambda_i})
{\Delta _x}({{\bf{R}}^\lambda_i}{\bf{ + r}})} \right\rangle }, 
\label{eq:Pd} 
\end{eqnarray}
where $\Delta_x^\dag ({{\bf{R}}^\lambda_i})$ is the creation operator of the NN singlet pair along the $x$-direction, defined as, 
\begin{eqnarray}
\Delta_x^\dag ({{\bf{R}}^\lambda_i}) = \frac{1}{\sqrt{2}}
(c_{i \uparrow}^{\lambda\dag} c_{i + x \downarrow}^{\lambda'\dag} 
+ c_{i + x \uparrow}^{\lambda'\dag} 
c_{i \downarrow}^{\lambda\dag}). 
\label{eq:DelR} 
\end{eqnarray}
When $P({\bf r})$ is finite in the limit of $|{\bf r}|\rightarrow\infty$, one can judge that an off-diagonal long-range order is realized. 
Instead of $P( |{\bf r}|\rightarrow\infty)$, we average $P({\bf r})$ for 
${\bf r}$'s, which are taken on a line segments of $(0,L/2)$-$(L,L/2)$ and 
$(L/2, 0)$-$(L/2,L)$ in the Type-II model with $L=12$.  
We note that $P({\bf r})$ is almost constant for $|{\bf r}| \ge 6$ in this system. 
In Fig.~\ref{fig:EcPd}(b), we present the averaged $P({\bf r})$, termed $P^{\rm ave}$, as a function of $\Delta$ at $\delta = 0.04$. 
It is shown that $P^{\rm ave}$ for the $d_1$-wave SC state is larger than that for the $d$+$id$-wave SC state, and has a large magnitude, in particular for $\Delta \le 2.5t$. 
These results indicate that the SC state of the $d_1$-wave gap is realized in the carrier-doped band insulator. 

\begin{figure}[t]
\begin{center}
\includegraphics[width=6.9cm,height=10.5cm]{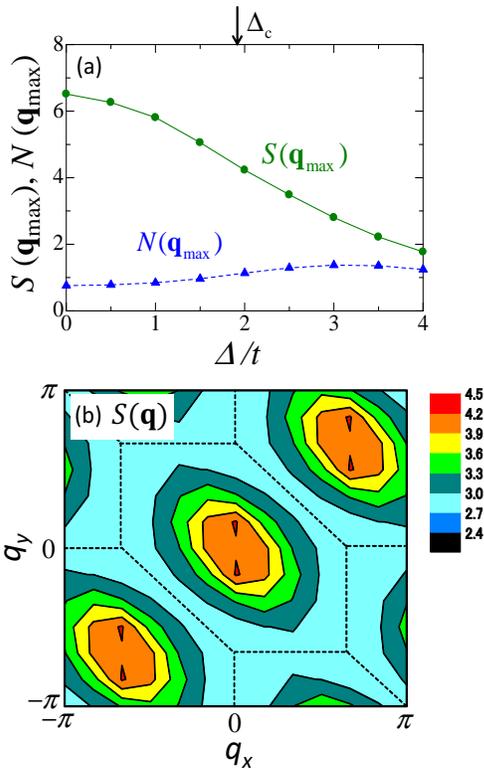}
\end{center}
\vspace{-0.5cm}
\caption{(Color online)
(a) The maxima of the spin and charge structure factors, $S({\bf q})$ and $N({\bf q})$, as functions of $\Delta/t$. 
We define ${\bf q}_{\rm max}$ to be the momenta at which the correlation functions take their maxima.  
The vertical arrow indicates $\Delta_{\rm c}$. 
(b) A contour plot of $S({\bf q})$ at $\Delta/t = 2$. 
The parameters are chosen to be $\delta = 0.04$, $t'= 0$, and $U/t =7$. 
The system size of $N_{\rm S} = 96$ is used in the Type II model. 
}
\label{fig:SqNq}
\end{figure}

In order to examine the origin of the SC state, we show in Fig.~\ref{fig:SqNq}(a) the maximum of the spin structure factor $S({\bf q})$ and the charge structure factor $N({\bf q})$ as functions of $\Delta$ for $\delta = 0.04$ and $t'/t=0$. 
The spin structure factor is defined as 
\begin{eqnarray}
S({\bf{q}}) &=& \frac{4}{{{N_{\rm{S}}}}}\sum\limits_{ij\lambda\lambda'} 
{{e^{i{\bf{q}} \cdot \left( {{{\bf{R}}^\lambda_i} - 
{{\bf{R}}^{\lambda'}_j}} \right)}}} \nonumber\\
&\times& \left( {\left\langle {S_i^{z\lambda}S_j^{z\lambda'}} \right\rangle 
- \left\langle {S_i^{z\lambda}} \right\rangle 
\left\langle {S_j^{z\lambda'}} \right\rangle } \right), 
\label{eq:Sq} 
\end{eqnarray}
and ${\bf q}_{\rm max}$ exhibits a wave number vector at which $S({\bf q})$ or $N({\bf q})$ takes their maxima. 
As shown in the figure, 
$S({\bf{q}}_{\rm max})$ has a large magnitude in the region of small $\Delta$, 
while $N({\bf{q}}_{\rm max})$ is almost unchanged by changing $\Delta$. 
The maxima of $S({\bf q})$ are located at ${\bf q}_{\rm max} = (0,\pm \pi/6)$ 
(see Fig.~\ref{fig:SqNq}(b)). 
Although $S({\bf{q}}_{\rm max})$ is rapidly reduced as $\Delta$ approaches $\Delta_{\rm c}$, it remains a  certain magnitude even in the region of $\Delta > \Delta_{\rm c}$. 
It is noticed that $S({\bf{q}}_{\rm max})$ shows a similar $\Delta$ dependence with $P^{\rm ave}$ shown in Fig.~\ref{fig:EcPd}(b). 
This fact suggests that the spin fluctuation is closely related to 
the $d_1$-wave SC correlation.

\begin{figure}[t]
\vspace{+0.9cm}
\begin{center}
\includegraphics[width=7.4cm,height=5.7cm]{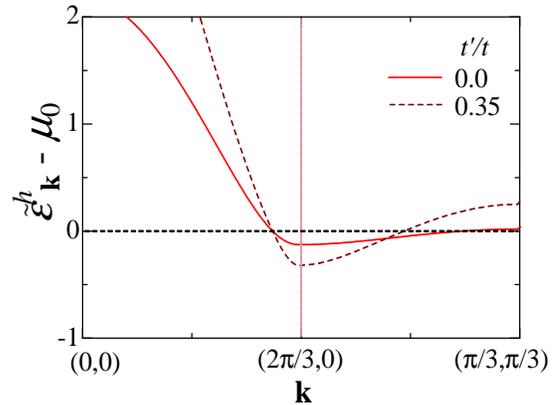}
\end{center}
\vspace{-0.5cm}
\caption{(Color online)
Energy bands $\tilde \varepsilon _{\bf k}^h - \mu_0$ 
for $t'/t=0$ and $0.35$ along 
${\bf k} = (0,0)$-$(2\pi/3,0)$-$(\pi/3,\pi/3)$. 
The parameter values are chosen to be $\Delta/t = 2$ and $U/t = 7$. 
The optimized value of $\Delta_{\rm CD}$ is used in 
$\tilde \varepsilon _{\bf k}^h$, 
and $\mu_0$ is determined from this band structure and $\delta =0.04$.  
}
\label{fig:comsu}
\end{figure}

Introduction of $t'$ causes a change in the above result for the SC state. 
For $t'/t = 0.35$, a phase diagram in the plane of $\Delta/t$ and $\delta$ 
is presented in Fig.~\ref{fig:dopeph}(b). 
The $d_1$- and $d$+$id$-wave states are stabilized only for $\delta \le 0.04$ 
and $\Delta \le 1.0$, and consequently the SC state does not appear 
in the region of the doped band insulator. 
To examine the effect of $t'$, in Fig.~\ref{fig:comsu}, we compare 
the energy bands for $t'/t=0$ and $0.35$ in the case of $\Delta/t = 2$. 
We plot $\tilde \varepsilon _{\bf k}^h$ introduced in Eq.~(\ref{eq:epk}), 
where the optimized $\Delta_{\rm CD}$ for $U/t = 7$ and $\mu_0$ for $\delta =0.04$ calculated in this band are adopted. 
When $t'/t=0$, the band is almost flat along $(2\pi/3, 0)$-
$(\pi/3, \pi/3)$, and its small curvature at 
${\bf k} = (2\pi/3,0)$ produces the large density of state at 
the Fermi energy. 
On the other hand, the curvature for $t'/t = 0.35$ is 
larger than that for $t'/t = 0$.
We interpret that the introduction of $t'$ decreases 
the density of state at the Fermi level and 
destabilizes the $d_1$-wave SC state. 

\section{Summary and discussion
\label{sec:summary}}

To elucidate the origin of SC in Li$_x$$M$NCl, we examined 
stability of the SC state in the ionic-Hubbard model on a single-layered 
honeycomb lattice, using the optimization VMC method. 
In the non-doped system, we investigated the $\Delta$-$U$ parameter space 
in which the band insulator is realized. 
For the parameter region corresponding to 
the carrier-doped band insulator, we studied the stabilities of 
the SC state with a variety of pairing symmetries. 
We found that, even in the carrier-doped band insulator, the spin fluctuation survives with sufficient magnitude. 
As shown in Fig.~\ref{fig:SqNq}(b), the spin correlation function $S({\bf q})$ 
takes its maximum around ${\bm q}=(0, \pm\pi/6)$. 
%
%
The $d_1$-wave type SC state, corresponding to the $d_{x^2-y^2}$-wave in the high-$T_{\rm c}$ cuprate, is stabilized near the half-filling, in comparison with the SC states with other symmetries including the $d+id$ wave.
This stability of the $d_1$-wave SC state might be related to the spin fluctuation maximized around ${\bm q}=(0, \pm\pi/6)$, because these wave vectors 
satisfy the nesting condition for one of the two Fermi pockets 
near the half filling. 
%
%

Our result supports some experimental observations in Li$_x$$M$NCl.\cite{Tou2,Hotehama,Kasahara} 
First, the SC phase appears in the electron-doped band insulator.
As shown in Fig.~\ref{fig:ph}, the correlation effect shifts the phase boundary toward the AF Mott insulating phase. 
That is to say, there is a parametr space for the band insulator which 
is stabilized by the quantum fluctuation from the Mott insulating phase. 
We suppose that a series of compound is located in such a region for the band insulator close to the phase boundary. 
The spin fluctuation is suppressed due to correlation effect in the band 
insulating state, but still survives.
We propose a possibility that remnant spin fluctuation in the doped band insualtor induces the superconductivity. 
Second point is about the SC state with unconventional spin singlet pair. 
The anisotropic gap was suggested by the NMR Knight-shift \cite{Tou2,Hotehama} 
and the specific heat measurements.\cite{Kasahara} 
From the present VMC calculations, one possibility for the gap symmetry is the $d_1$-wave shown in Fig.~\ref{fig:gap} which is stabilized in the lightly doped band insulator close to the phase boundary with the Mott insulator (see Fig.~\ref{fig:dopeph}). 
This is related to the spin fluctuation around ${\bm q}=(0, \pm\pi/6)$, 
as mentioned above. 
Some questions caused from the experimental observations in Li$_x$$M$NCl 
are not resolved within the present theoretical analyses.  
One is the doping concentration dependence of $T_{\rm c}$, which is nearly constant ($\sim 10$K) in a region of $\delta > 0.10$,\cite{Kasahara,Ye} although the SC is stabilized only for $\delta \lsim 0.04$ in our result. 
To solve this question, further factors to stabilize the SC might be required. 
Charge fluctuations due to the inter-site Coulomb interactions and the interlayer interactions in the conducting plane might be candidates. 


\begin{acknowledgments}
The authors appreciate Y. Iwasa, and Y. Taguchi for their valuable discussion. This work is supported by KAKENHI 
from MEXT. 
\end{acknowledgments} 
\par




\begin{thebibliography}{99}
\bibitem{pair} Although SC pairing terms for intra-band and inter-band pairings appear, the inter-band pairing term which is unimportant at zero temperature is dropped.
\end{thebibliography}
\end{document}